\documentclass[11pt,a4paper]{article}

\usepackage{authblk}

\usepackage[top=45mm, bottom=45mm, left=35.35mm, right=35.35mm]{geometry}
\usepackage{eucal}
\usepackage[sfdefault]{FiraSans}
\usepackage{nth}
\usepackage{parskip}
\RequirePackage[colorlinks,citecolor=blue,urlcolor=blue,linkcolor=blue]{hyperref}

\usepackage{amsthm,amsmath,amsfonts,amssymb}
\usepackage{mathtools}
\usepackage{knockoffs-paper-macros}

\usepackage[authoryear,round]{natbib}

\usepackage{multirow}
\usepackage{booktabs}
\usepackage{tabularx}
\usepackage{graphicx}
\usepackage[usenames,dvipsnames,table]{xcolor}
\usepackage{subcaption}
\graphicspath{{figs/}}

\usepackage{algorithm,algpseudocode}
\algrenewcommand\algorithmicindent{1em}
\algnewcommand\algorithmicinput{\textbf{Input:}}
\algnewcommand\algorithmicoutput{\textbf{Output:}}
\algnewcommand\Input{\item[\algorithmicinput]}%
\algnewcommand\Output{\item[\algorithmicoutput]}%

\usepackage[skins]{tcolorbox}
\newcommand{\luke}[1]{\footnote{\textcolor{Orange}{Luke: #1}}}

\title{Sequential knockoffs for continuous and categorical predictors: with application to a large Psoriatic Arthritis clinical trial pool}
\author[1]{Matthias Kormaksson}
\author[2]{Luke J. Kelly}
\author[1]{Xuan Zhu}
\author[1]{Sibylle Haemmerle}
\author[1]{Luminita Pricop}
\author[1]{David Ohlssen}

\affil[1]{Novartis Pharmaceuticals Corporation}
\affil[2]{Oxford University}

\begin{document}

\maketitle

\begin{abstract}

Knockoffs provide a general framework for controlling the false discovery rate when performing variable selection. Much of the Knockoffs literature focuses on theoretical challenges and we recognize a need for bringing some of the current ideas into practice. In this paper we propose a sequential algorithm for generating knockoffs when underlying data consists of both continuous and categorical (factor) variables. Further, we present a heuristic multiple knockoffs approach that offers a practical assessment of how robust the knockoff selection process is for a given data set. We conduct extensive simulations to validate performance of the proposed methodology. Finally, we demonstrate the utility of the methods on a large clinical data pool of more than $2,000$ patients with psoriatic arthritis evaluated in 4 clinical trials with an IL-17A inhibitor, secukinumab (Cosentyx), where we determine prognostic factors of a well established clinical outcome. The analyses presented in this paper could provide a wide range of applications to commonly encountered data sets in medical practice and other fields where variable selection is of particular interest.
\end{abstract}


\section{Introduction}

A successful drug development program will often lead to a substantial amount of data from multiple clinical trials at different phases I-IV. These individual trial data can be synthesized and modeled to answer questions
beyond the intended primary purpose of any one study. For example,
a larger pool of data could be used to examine a greater level of
understanding of factors affecting either treatment effect heterogeneity
or long term prognosis of a patient. The search for such prognostic or predictive factors may be framed as a variable selection problem and mathematically, this could be expressed as understanding the conditional distribution of an outcome $Y$ given a set of covariates $X$. It is this setting that will form the focus of work here.
%


Classical hypothesis testing leading to p-values has provided a cornerstone
in determining success in individual clinical trials, where pre-specified tests aim at determining safety, optimal dose, or efficacy of a given drug. However, the problem of understanding which covariates (among many) are driving prognosis is very different to the pre-specified analysis of a single trial and either test statistics or p-values with well understood mathematical properties are more
difficult to calculate. These challenges, and the lack of statistical rigor in addressing them, have led to the so-called replication crisis \citep{wasserstein}, i.e. the problem of false (non-reproducible) discoveries in scientific publications.

The seminal paper of \citet{benjamini95} introduced a method for controlling the False Discovery Rate (FDR), the expected proportion of false discoveries among the variables selected. With the advent of (big) microarray data, similar statistical ideas and extensions became increasingly popular (e.g. \cite{limma} and \cite{lfdr}). In short, these methods involve producing test statistics or p-values measuring association between covariates and response. Typically, the set of tests would then be combined with a method to control multiple comparisons or borrow information across tests. \cite{benjamini01} discusses the conditions on when Benjamini--Hochberg works and refers to a simulation paper where they showed the FDR was controlled in more general situations than independent p-values.

Parametric regression modeling is perhaps the most widely utilized
approach to assess conditional probability relationships,
particularly in epidemiological analyses. However, although results
from fitting such models are often interpreted in terms of conditional
relationships (by using confidence intervals or p-values associated
with model parameters) the interpretation is rather unclear due to
the need to account for multiple comparisons. \cite{harrell} provides general problem solving strategies for model selection in regression, but many of such strategies are difficult to automate and focus on inference as opposed to variable selection and false discovery rate control.

\cite{hastie} compiled a vast list of material on modeling techniques both from classical statistics and the machine learning literature. Most of these methods tend to focus on forming a good model for prediction, while others, such as lasso or elastic net \citep{tibshirani1996, zou05} also provide a final set of selected variables. These methods have gained widespread popularity since their introduction to the literature, in particular due to the fact that they can handle a large number of explanatory variables. However, some recent work \citep{su2017} has observed that lasso has problems in selecting the proper model in practical applications, and that false discoveries may appear very early on the lasso path. Finally, it's worth pointing out that until recently there were no particular theoretical results associated with the appropriateness of the lasso selection, e.g. via p-values or confidence intervals. However, recent publications on post-selection inference \citep{lee16,tibshirani16} and sure independence screening \citep{fan} provide steps in this direction and remains an exciting and active research area. 

The review underlines that the approaches commonly used in practice
all have limitations. Therefore, it is natural to look at the
mathematical statistics literature for developments that have the
potential to move into practice. One such development is the so called
\emph{knockoff} approach \citep{barber15,candes18,romano19,sesia19}. The method directly addresses the conditional relationship question and allows one to identify the truly important predictors, while controlling the FDR. The method has been implemented into an R-package \citep{knockoffpackage}, which may be used in applications. However, this implementation only considers the case where all covariates are adequately represented by the Gaussian distribution. In this paper we present a novel practical algorithm, inspired by an algorithmic blueprint of \cite{candes18}, that may be used to analyze more general data sets involving a mixture of continuous and categorical explanatory variables.

Another practical limitation of the knockoff approach is that simulating a single knockoff may result in different variable selections if run twice. Recent work \citep{holden18, gimenez19, nguyen20} offer some alternative solutions involving multiple knockoffs with FDR control. However, to the best of our knowledge there is no computational software package available for generating such multiple knockoffs. Therefore, as a second novelty of this paper, we propose a simple and heuristic method for selecting influential variables using multiple knockoffs of the same data set.

In the remainder of this paper we will examine the knockoff approach
in the context of identifying prognostic factors related to a clinical
outcome in a large clinical trial pool. In \secref{sec:method} we describe the knockoff methodology and develop some extensions to deal with combinations of mixed covariate data (e.g. discrete and continuous) and the possibility of using multiple knockoffs to assess stability. \secref{sec:simulations} then provides an extensive simulation study comparing the knockoff methodology against a number of competitors in settings similar to the application. In \secref{sec:application} we perform a detailed analysis of a large psoriatic arthritis clinical trial database, which demonstrates the utility of the methods. Finally, \secref{sec:conclusion} summarizes our main findings and provides a discussion on possible areas for further research.


\section{Method}
\label{sec:method}

We consider the problem of variable selection in the context of clinical data $(Y, X)$, where $Y$ denotes a clinical outcome of interest and $X = (X_1, \dotsc, X_p)$ denotes patient measurements (e.g. baseline variables at randomization), which may be continuous or categorical. Typical applications include generalized linear models for continuous or categorical outcomes and Cox regression for time to event endpoints.

Mathematically, we may describe the problem as follows: Let $ S \subseteq \{1, \dots p\} $ denote the indices of the variables in $ X = (X_1, \dotsc, X_p) $ which influence the outcome $ Y $ given the other measurements. In other words, we seek the smallest set $ S $ such that
\[
    Y \indep X_{S'} \given X_S,
\]
where the complement set $ S' = \{1,\dots, p\} \setminus S $ indexes the \emph{null} variables for the problem. Any variable selection procedure $ \hat{S} $ will estimate $ S $ with error. For our purposes, we seek an estimator $ \hat{S} $ of $ S $ which controls the false discovery rate (FDR), the expected false discovery proportion (FDP), or the expected proportion of null variables returned by $ \hat{S} $,
\begin{equation}
    \label{eq:fdr}
    \mathsf{FDR} = \EE[\mathsf{FDP}(\hat{S})] = \EE\left[\frac{\abs{\hat{S} \setminus S}}{\abs{\hat{S}} \vee 1}\right],
\end{equation}
while also having the statistical power to detect associations. Note that the expectation in \eqnref{eq:fdr} is with respect to the data-generating distribution: the FDP for any realisation of an FDR-controlling procedure will vary around the FDR due to randomness in the data and, possibly, the selection procedure $ \hat{S} $.

\subsection{Background} \label{subsec:background}

\citet{barber15} proposed the \emph{knockoff filter} to select variables controlling the FDR with finite sample guarantees. We first construct \emph{knockoffs} $ \tilde{X} $ for the observed covariates $ X $ then compare their ability to model the response $ Y $. A subsequent selection step returns the variables which sufficiently outperform their knockoffs while also controlling the FDR. \citet{barber15} construct a fixed knockoff matrix $ \tilde{X} $ which mimics the correlation structure within $ X $ but minimises the cross-correlation between $ X $ and $ \tilde{X} $. 

\citet{candes18} proposed the Model-X knockoff framework, a more flexible approach valid regardless of the distribution of $ Y \given X $, where we build an appropriate coupling of $ P_X $ and $ P_{\tilde{X}} $ then sample knockoffs from the conditional distribution $ \tilde{X} \given X $. In the Model-X framework, a knockoff $ \tilde{X} $ for $ X $ satisfies the following properties:
\newpage
\begin{enumerate}
    \item[(P1)] $ \tilde{X} \indep Y \given X$,
    \item[(P2)] $ [X, \tilde{X}] \stackrel{d}{=} [X, \tilde{X}]_{\mathsf{swap}(A)} $ for any $ A \subseteq S' $,
\end{enumerate}
where $\mathsf{swap}(A)$ applied to $[X, \tilde{X}]$ exchanges $X_j$ and $\tilde{X}_j$ for each $j \in A$. P1 is trivially satisfied by sampling knockoffs without looking at $ Y $, while P2 requires that swapping null variables and their knockoffs does not affect the joint distribution of $ X $ and $ \tilde{X} $.

With properties P1 and P2 satisfied, we can use knockoffs to guarantee FDR control at a given level $ q \in (0, 1) $:
\begin{enumerate}
    \item For each variable $ j $, compute a feature statistic $ W_j = w_j((X, \tilde{X}), Y) $ to distinguish between $ X_j $ and its knockoff $ \tilde{X}_j $ (for example, $W_j = |\beta_j|-|\tilde{\beta}_j|$ in an augmented regression of $Y$ against $X$ and $\tilde{X}$); large, positive statistics indicate association with $ Y $.
    \item Use the \emph{knockoffs$+$} procedure to select variables: return $ \hat{S} = \{j : W_j \geq \tau_+\} $ where
    \begin{equation}
        \label{eq:k+}
        \tau_+ = \argmin_{t > 0} \left\{\frac{1 + \abs{\{j : W_j \leq t\}}}{\abs{\{j : W_j \leq t\}}} \leq q\right\}.
    \end{equation}
\end{enumerate}

\citet{candes18} describe how to sample Model-X knockoffs in the Gaussian case by directly sampling from the conditional distribution of $ \tilde{X} \given X $. That is, if $ X \sim \cN(0, \bSigma) $, then
\[
(X, \tilde{X}) \sim \cN(0, \bG), \quad \text{where} \quad \bG = \begin{pmatrix} \bSigma & \bSigma - \diag(s) \\ \bSigma - \diag(s) & \bSigma \end{pmatrix},
\]
and $ s $ is chosen such that $ \bG \succeq 0 $; we sample $ \tilde{X} \given X \sim \cN(\mu, \bV) $ where
\begin{align*}
	\mu &= X \big(\bI - \bSigma^{-1} \diag(s)\big), \\
	\bV &= 2 \diag(s) - \diag(s) \bSigma^{-1} \diag(s).
\end{align*}
This approach requires that $ X $ is adequately modelled by a Gaussian distribution. 

Building on the Model-X approach, \citet{romano19} describe how to generate Model-X knockoffs using deep neural nets, \citet{sesia19} develop hidden Markov model knockoffs for sequence data, and \citet{barber19} propose a knockoff filter for high-dimensional data by incorporating a screening step. Both the Gaussian Model-X and hidden Markov model approaches are special cases of a more general recipe, Sequential Conditional Independent Pairs, also described by \citet{candes18}, which we reproduce in \algoref{alg:scip} and which forms the basis for our procedure.

\begin{myalgorithm}[Sequential conditional independent pairs]
    \label{alg:scip}
    For $ j = 1, \dotsc, p $, sample
    \[
        \tilde{X}_j \sim \cL(X_j \given X_{-j}, \tilde{X}_{1:j-1}),
    \]
    where $ X_{-j} \defeq (X_1, \dotsc, X_{j-1}, X_{j+1}, \dotsc, X_p) $ and $ \tilde{X}_{1:j-1} \defeq (\tilde{X}_1, \dotsc, \tilde{X}_{j-1}) $.
\end{myalgorithm}

\subsection{Sequential knockoff algorithm for mixed data types} \label{subsec:seqknockoff}

We are interested in generating knockoffs for mixed data types coming from clinical trials, so it does not fall into the purely continuous or discrete approaches in the literature \citep{candes18,romano19,sesia19}. In the following, we describe a brute-force \emph{sequential knockoff algorithm} where we estimate and sample from the conditional distributions in \algoref{alg:scip}.

\begin{myalgorithm}[Sequential knockoff algorithm for mixed data types] \label{alg:scip-approx}

Recall definitions from \algoref{alg:scip}. We use elastic net penalised generalized linear models at each step to estimate the parameters of the sequential conditional distributions.
More specifically, for $ j = 1, \dotsc, p $,
    \begin{itemize}
    	\item if $X_j$ is continuous, fit a penalized linear regression model with response $X_j$ and covariates $X_{-j}$ and $\tilde{X}_{1:j-1}$.

    	Sample $\tilde{X}_j \sim \cN\left(\hat{\mu}, \hat{\sigma}^2\right)$, where $\hat{\mu}$ and $\hat{\sigma}^2$ are estimates of the regression mean and error variance, respectively.
    	\item if $X_j$ is categorical, fit a penalized multinomial logistic regression model with response $X_j$ and covariates $X_{-j}$ and $\tilde{X}_{1:j-1}$.

    	Sample $\tilde{X}_j \sim {\mathrm{Multinom}}(\hat{\pi})$, where $\hat{\pi}$ denotes the regression estimate of the multinomial probabilities.
    \end{itemize}
    Return $ \tilde{X} = (\tilde{X}_1, \dotsc, \tilde{X}_p) $.
\end{myalgorithm}

When we say ``fit model'' above, it is to be understood in the context of data consisting of $n$ i.i.d. samples of $(X, Y)$. We validate our approach in \secref{sec:simulations} and describe the finer details of the procedure in \appref{app:senk}.

\textit{Remark:} For theoretical FDR control, \algoref{alg:scip} (and Model-X knockoffs in general) require (oracle) knowledge of the underlying distributions \citep{candes18} or sufficient data to estimate them accurately \citep{romano19,sesia19}. For that reason \cite{barber20} investigate robustness of Model-X knockoffs to model misspecification and establish a bound for the error in controlling FDR in terms of the Kullback-Leibler divergence between exact and approximate conditional distributions. However, although our simulations do not yield a useful bound in terms of KL divergences, they demonstrate that the sequential knockoff procedure (\algoref{alg:scip-approx}) empirically controls the FDR.

\subsection{Variable selection in practice using multiple knockoffs} \label{subsec:multiknockoff}

In practice, simulating a single knockoff $\tilde{X}$ and reporting the resulting variables may not be compelling. Particularly when the quantitative scientist conveys to the clinical investigator that the selection in fact varies and depends on the random seed of the algorithm. Recent work \citep{holden18, gimenez19, nguyen20} offer some alternative solutions involving multiple knockoffs with FDR control. However, to the best of our knowledge there is no computational software package available for generating such multiple knockoffs.

Therefore, we propose a heuristic method for selecting influential variables using multiple knockoffs of the same data set. Although we do not offer theoretical properties of this approach, we shall examine it with extensive simulations in \secref{sec:simulations}. The proposed method gives practitioners, in addition to the selected variables, a sense of how much uncertainty there is in the selection procedure.

Let $\tilde{X}_1, \dots, \tilde{X}_B$ denote $B$ independent knockoff copies of $X$ (e.g. generated with \algoref{alg:scip-approx}). For each knockoff copy $b$ we run the knockoff filter, outlined in \secref{subsec:background}, and select the set of influential variables, $S_b \subseteq \{1,\dots,p\}$. Let $I_{jb}=1\{j \in S_b\}$ denote the indicator for whether variable $j$ was selected in knockoff draw $b$. The matrix $(I_{jb})$ may be used to visualize (with a heatmap) the robustness and uncertainty in the knockoff variable selection process (see \figref{fig:heatmap_acr20} for an example heatmap). If there is a lot of noise in the heatmap of $(I_{jb})$ then we may not find the knockoff selection trustworthy. To formalize the concept of a ``trustworthy'' selection we propose the following heuristic:
\begin{enumerate}
	\item Let $F(r) \subseteq \{1,\dots,p\}$, where $r \in [0.5, 1]$, denote the set of variables selected more than $r \cdot B$ times out of the $B$ knockoff draws.
	\item Let $S(r)=\underset{b}{\rm mode}\{F(r) \cap S_b\}$ denote the set of selected variables that appears most frequently, after filtering out variables that are not in $F(r)$.
	\item Return $ \hat{S} = S(\hat{r}) $, where $ \hat{r} = \argmax_{r \geq 0.5} \abs{S(r)}$, i.e. the largest set among $\{S(r):r \geq 0.5\}$.
\end{enumerate}
The first step above essentially filters out variables that don't appear more than $(100\cdot r)\%$ of the time, which would seem like a reasonable requirement in practice (e.g. with $r=0.5$). The second step above then filters the $B$ knockoff selections $S_b$ accordingly and searches for the most frequent variable set among those. The third step then establishes the final selection, namely the most liberal variable selection among the sets $\{S(r):r \geq 0.5\}$.

\textit{Remark A:} Filtering prior to searching for most frequent ``model'' turns out to be important in practice as spurious inclusions/exclusions of marginally significant variables may lead to the non-existence of a single most frequent model (e.g. each set $S_b$ could be unique despite sharing a core set of common variables).

\textit{Remark B:} Performing step 3 above may seem unnecessary at first glance; one might e.g. instead simply choose $S = S(0.5)$. However, it's not hard to construct an example where $S(0.5) = \emptyset$, while $S(r) \neq \emptyset$ for some $r > 0.5$. In those types of scenarios the latter non-empty set would be a sensible (and preferred) selection.

\textit{Remark C:} It is worth re-emphasizing that the proposed approach is heuristic and care must be taken when interpreting and processing the heatmap of the multiple knockoff selections $(I_{jb})$. An example where the above multiple selection procedure could go wrong is when there are high correlations among predictors entering the model. In this case the most important variables may be correlated so only one is picked per knockoff data set, due to sampling variability, while the less important variables may be independent and selected all of the time. However, such issues can be avoided by carefully addressing multicollinearity issues prior to modeling.

\subsection{Competing methods} \label{subsec:competingmethods}

Here, we briefly review some popular alternative approaches that will be used as benchmarks in our simulation experiments in \secref{sec:simulations}:

\emph{Adjusted p-values:} As a first comparator we consider adjusted p-values \citep{benjamini95} from partial slope t-tests of multiple regression. We also consider the \cite{benjamini01} extension for dependent p-values.

\emph{Permutation LASSO:} \cite{permlasso2} provide a review of some approaches aimed at controlling FDR in genome-wide association studies and among those we choose Permutation LASSO as our second comparator. In short, we calculate the LASSO path both on the original data and $B$ random permutations, on the same $\lambda$-grid. We then choose $\lambda$ that controls a slightly modified version of FDR, with numerator given by the average number of (false) discoveries from the permutated data, and denominator the number of LASSO selections from the original data. 

\emph{Post-selection inference:} \cite{tibshirani16} proposed new inference tools for Forward Stepwise Regression (FSR), Least Angle Regression (LAR), and the LASSO. In particular, they advocate the idea of computing p-values across steps of the regression (FSR, LAR, or LASSO) and then apply the sequential ``Forward Stop'' rule of \cite{gsell}, which guarantees FDR control. 

\subsection{Computations}

The sequential and multiple knockoffs methods from Sections \ref{subsec:seqknockoff} and \ref{subsec:multiknockoff} we have implemented in an R-package available at \url{https://github.com/kormama1/seqknockoff}. For Model-X knockoffs and the competing methods in \secref{subsec:competingmethods} we in addition to the R base package used \emph{knockoff} \citep{knockoffpackage}, \emph{glmnet} \citep{glmnet}, and \emph{selectiveInference} \citep{selectiveInference}.

\section{Simulations}
\label{sec:simulations}

In this section we perform extensive simulations and compare performance of the different methods presented in \secref{sec:method}. 

\subsection{Description of simulated data sets and experimental parameters} \label{subsec:simdescription}

Below we describe in detail the simulation of a single data set $(X, y)$ for a given parameter configuration. For each parameter configuration we simulate $n_{sim}=100$ data sets and apply the different variable selection algorithms.

\emph{Simulation of $X$:} We simulate the rows of the $n \times p$ design matrix $X$ independently from a multivariate Gaussian distribution with mean $0$ and $p \times p$ covariance matrix $\Sigma = (\Sigma_{ij})$. Similar to \citet{candes18}, to cover a wide range of correlation structures we consider covariance matrices of the form:
\begin{equation}
\Sigma_{ij} = \left \{
\begin{array}{lr}
1\{i \neq j\}/n, & \text{Independent,} \\
\rho^{1\{i \neq j\}}/n, & \text{Equicorrelated,} \\
\rho^{|i-j|}/n, & \text{AR1}.
\end{array}
\right.
\end{equation}
We then randomly select $p_b$ of the columns of $X$ and dichotomize with the indicator function $\delta(x)=1(x > 0)$. The binarized columns are scaled by a $\sqrt{n}$-factor to ensure that the marginal (column-wise) variances are equal. The case $0 < p_b < p$ emulates a typical mixed-type clinical data application where some variables are continuous while others are categorical.

\emph{Simulation of $y | X$:} We simulate the response vector $y$ from a sparse Gaussian regression model $y \sim N(X\beta, I_n)$, where $p_{nn}$ of the $\beta$-components (corresponding to non-null features) have amplitude $a$, while all others are set to zero. The indices of the non-null features are sampled at random for each generated data set.

A summary of the underlying simulation parameters can be found in \tabref{tab:parameters}.

\begin{table}[tb]
	\centering
	\caption{Experimental parameters for simulation experiments.}
	\label{tab:parameters}
	\begin{tabular}{@{}rl@{}}
		\toprule
		$n$ & Number of data observations  \\
		$p$ & Number of covariates   \\
		$p_b$ & Number of binarized covariates   \\
		$\rho$ & Correlation coefficient  \\
		$p_{nn}$ & Number of non-null features  \\
		$a$ & Regression coefficient amplitude \\
		\bottomrule
	\end{tabular}
\end{table}

\subsection{Simulation study to assess performance of sequential knockoff filter} \label{subsec:simstudy1}


\begin{figure}[tb]
	\centering
	\includegraphics[width=\textwidth]{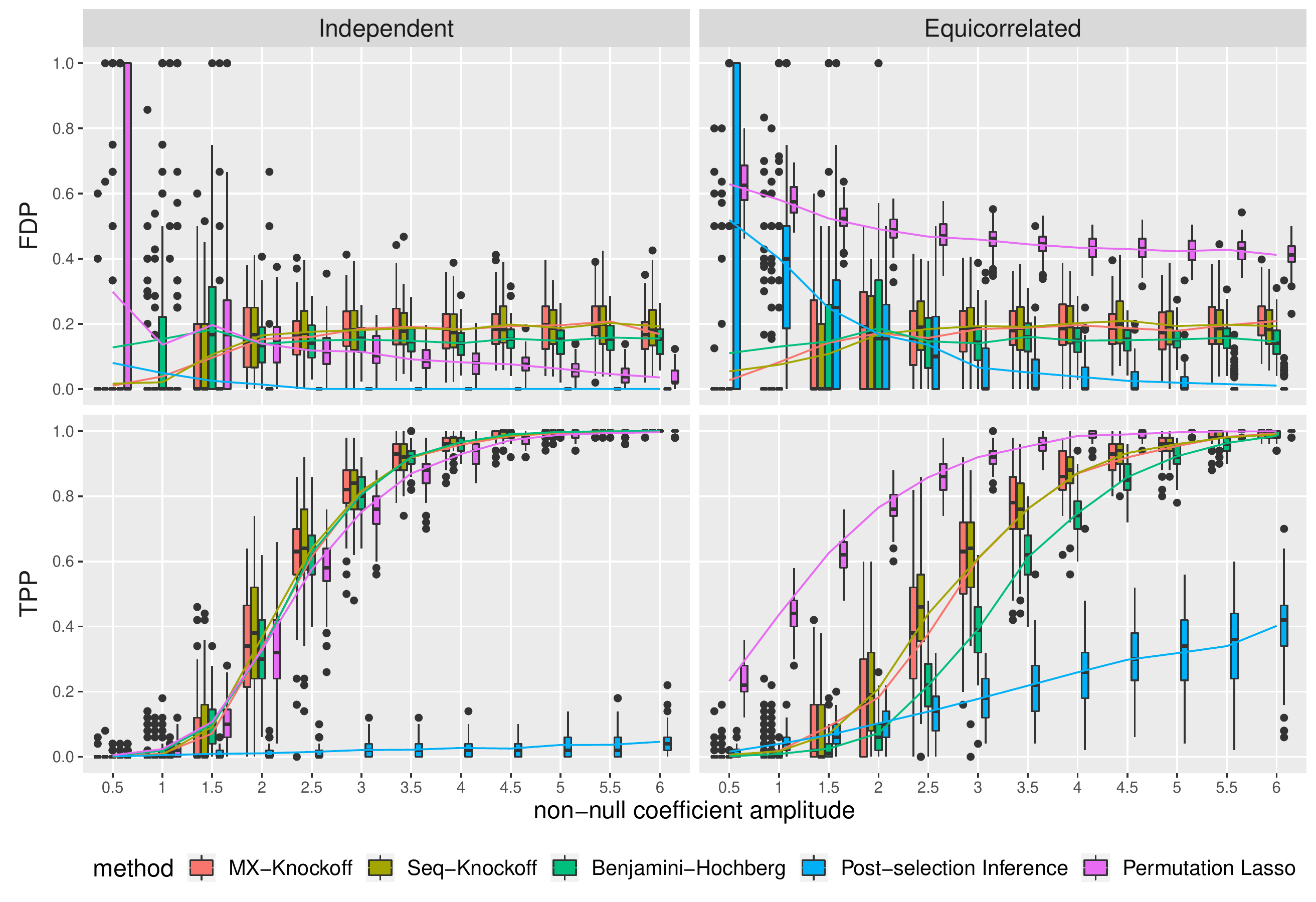}
	\caption{Comparing performance as the effect amplitude $ a $ varies. Left and right panels show results from independent and equicorrelated Gaussian design matrices, respectively. The colored mean curves provide estimates of FDR and power.}
	\label{fig:varying_a}
\end{figure}

\subsubsection{Varying amplitude} \label{subsec:varying_a}

In this experiment we vary the amplitude $a \in \{0.5, 1, \dots, 6\}$ and consider the continuous design matrix setting ($p_b=0$) with $n = 2,000$, $p=200$, $\rho=0.5$, and $p_{nn}=50$. In \figref{fig:varying_a} we compare the distributions of observed False Discovery Proportions (FDP) and True Positive Proportions (TPP) across the five competing methods. We note that the sequential knockoff method performs similarly to the MX knockoff approach in all scenarios; a phenomenon that we observed consistently in all other experimental settings involving a purely Gaussian design matrix. We further note that all methods controlled FDR in the independent covariate setting, while only the knockoff approach and Benjamini--Hochberg (BH) successfully controlled it in the equicorrelated setting (target threshold was $q=0.2$). In terms of power the BH method performed similar to the knockoff approach in the independent setting, but had slightly lower power in the equicorrelated setting. Similar results (to the equicorrelated case) were found in the AR1 covariance setting (not shown). The post-selection inference method, which is not purely geared towards variable selection, did not have competitive power in any of the scenarios and the permutation lasso method substantially failed to control FDR in the equicorrelated case. The Benjamini--Yekutieli method (not shown) controlled FDR in all scenarios, but had uniformly lower power than BH.

\begin{figure}[tb]
	\centering
	\includegraphics[width=\textwidth]{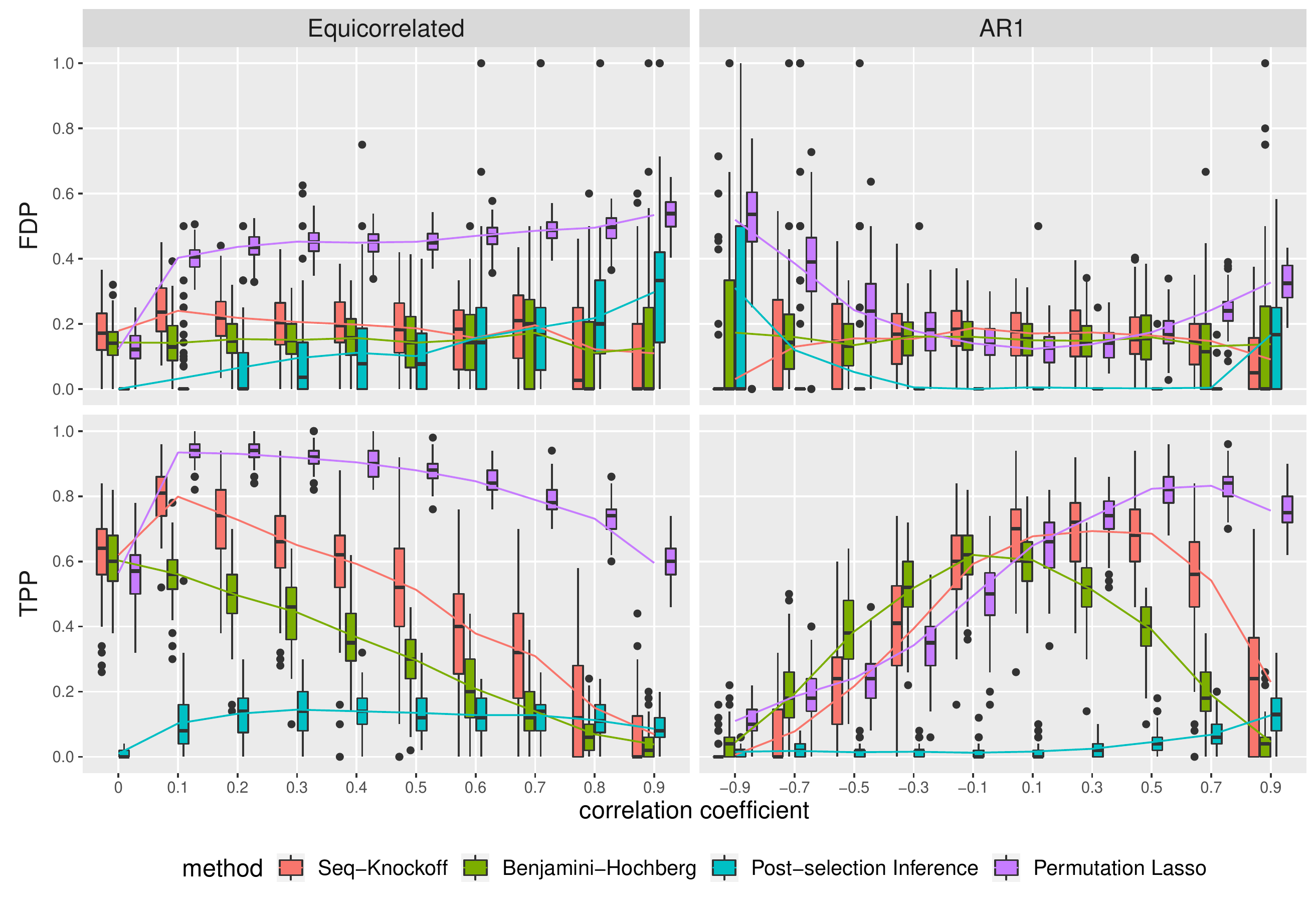}
	\caption{Comparing performance as $ \rho $ varies. Left and right panels respectively show results from equicorrelated and AR1 design matrices (with half of columns dichotomized). The colored mean curves provide estimates of FDR and power.}
	\label{fig:varying_rho}
\end{figure}

\subsubsection{Varying correlation coefficient} \label{subsec:varying_rho}

In this experiment we vary the correlation coefficient $\rho$ in the equicorrelated and AR1 covariance setting. We specifically consider the mixed-type variable setting with $p_b=100$ binarized covariates out of $p=200$. Additionally we set $n = 2,000$, $p_{nn}=50$, and $a=2.5$. In \figref{fig:varying_rho} we see how the FDP and TPP distributions vary across the methods as a function of $\rho$ in the two covariance settings. We note that the sequential knockoff algorithm and BH method control FDR (target threshold $q=0.2$) fairly consistently across the different correlation structures. Conversely, the permutation lasso (in most scenarios) and post-selection inference (in more extreme correlation scenarios) fail to control FDR. In terms of power, the knockoff method outperforms BH, but both deteriorate as inter-variable correlations become more extreme. A curious scenario occurs in the AR1 setting with $\rho < 0$ where BH outperforms the knockoff method in terms of power. However, this scenario is somewhat unrealistic in practice as it simulates several negatively correlated pairs of variables $(X_j, X_{j'})$ that are then both positively associated with $Y$. Finally, we re-emphasize that the R-implementation of MX knockoff \citep{knockoffpackage} only handles continuous design matrices and hence cannot be applied in this mixed-type covariate setting.

\begin{figure}[tb]
	\centering
	\includegraphics[width=\textwidth]{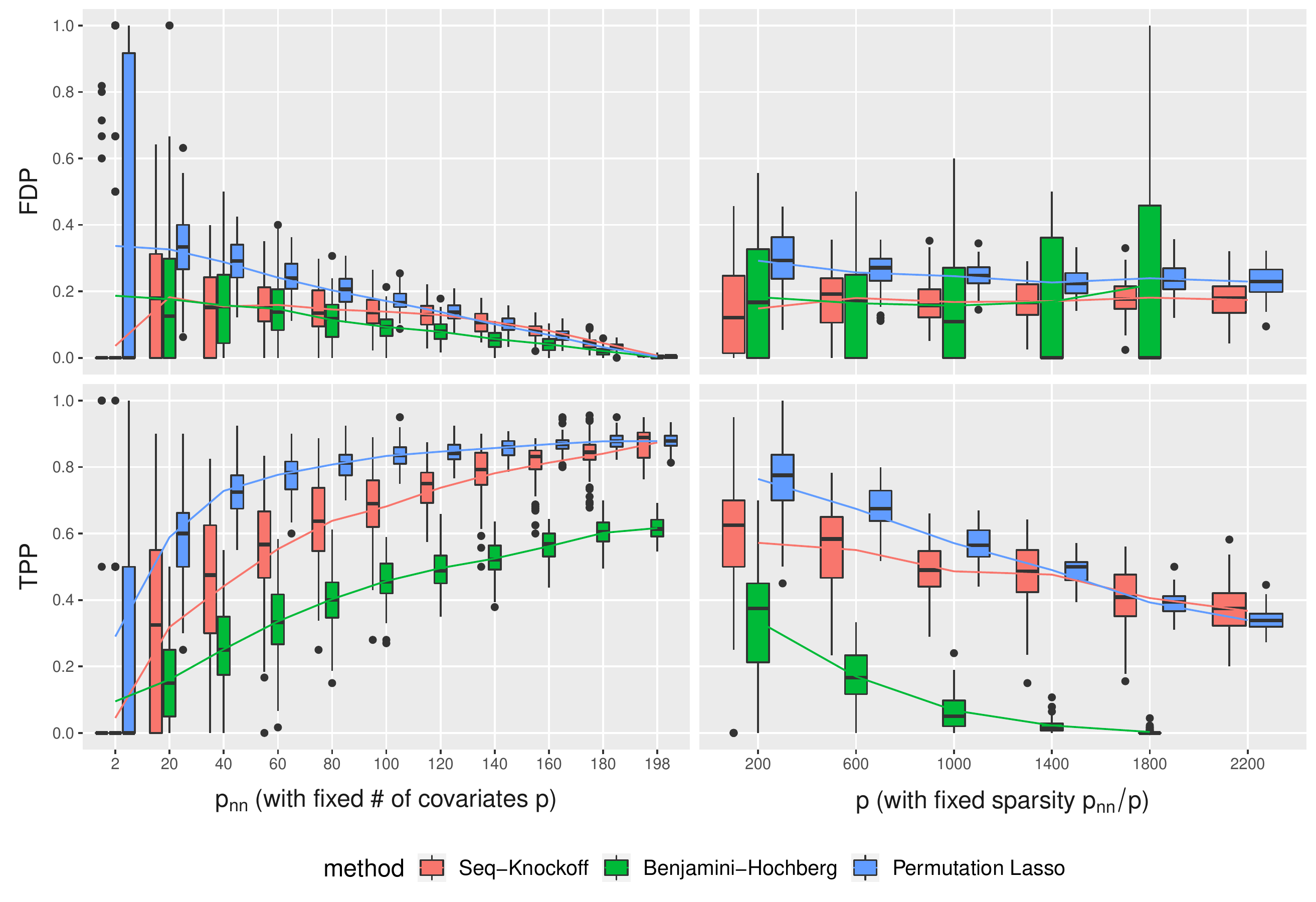}
	\caption{Left panel compares performance as the number of non-null features $p_{nn}$ increases (with $p=200$). Right panel compares performance as $p$ increases (with $p_{nn}/p = 0.1$). The colored mean curves provide estimates of FDR and power.}
	\label{fig:varying_p}
\end{figure}

\subsubsection{Varying sparsity and increased number of covariates} \label{subsec:varying_p}

In this experiment we compare two parameter settings:
\begin{enumerate}
	\item varying $p_{nn}$ with a fixed number of covariates $p=200$. \label{item:vary_p_nn}
	\item varying $p$ with a fixed sparsity $p_{nn}/p = 0.1$. \label{item:vary_p}
\end{enumerate}
Other parameters are set to $n = 2,000$, $p_b=0$, $\rho=0.5$, $a=2.5$ and we focus on the AR1 covariance setting. In \figref{fig:varying_p} we see how the FDP and TPP distributions vary as a function of $p_{nn}$ and $p$ under these two scenarios (left and right panel, respectively). We note that the sequential knockoff algorithm and BH both control FDR in all scenarios (target threshold $q=0.2$) while the former has higher power overall. Again permutation lasso fails to control FDR (in particular when the signal is sparse, i.e. when $p_{nn}/p$ is small). It is worth noting that as $p$ increases the benefits of the knockoff approach as compared to BH become evident. While the power of BH drops to zero as $p$ approaches the sample size $n=2,000$, the sequential knockoff algorithm only gradually loses power (even as $p$ increases beyond $n$).

\begin{figure}[tb]
	\centering
	\includegraphics[width=\textwidth]{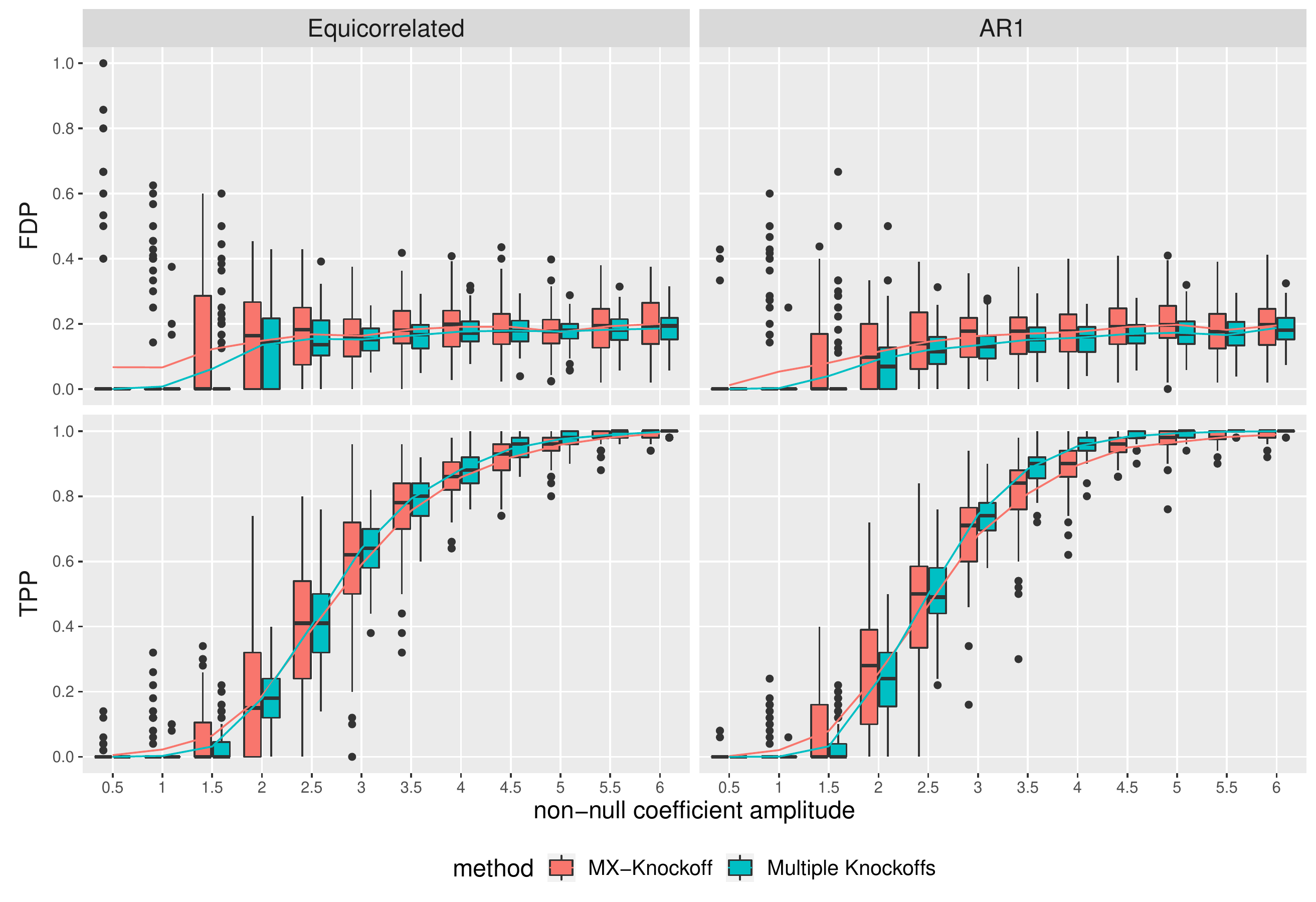}
	\caption{Comparing performance of the single and multiple knockoffs algorithm. Left and right panels show results from equicorrelated and autoregressive Gaussian design matrices, respectively. The colored mean curves provide estimates of FDR and power.}
	\label{fig:multiknock}
\end{figure}

\subsection{Multiple knockoffs filter empirically controls FDR} \label{subsec:multiknock}

In this section we compare the performance of the proposed multiple knockoffs filter from section \ref{subsec:multiknockoff} (with $1,000$ knockoffs per simulated data set) and the classical (single) knockoff algorithm. To that end, we perform a simulation study that shares the same experimental parameters as the "varying amplitude" simulation of \secref{subsec:varying_a} ($a \in \{0.5, 1, \dots, 6\}$, $n = 2,000$, $p=200$, $p_b=0$, $\rho=0.5$, and $p_{nn}=50$). Since in the ``varying amplitude'' experiment we established that our sequential knockoff and MX knockoff \citep{knockoffpackage} behave similarly across all simulation scenarios, we applied MX knockoff as our default knockoff generator in both the single and multiple knockoffs algorithm to gain computational speed\footnote{In this experiment we are simulating in total $3,600$ data sets ($12$ amplitudes, $3$ covariance structures, and $100$ data sets per parameter configuration). Further, the multiple knockoff algorithm involves simulating $1,000$ knockoffs per data set. We discuss computational aspects further in our conclusions.}.

In \figref{fig:multiknock} we see the results from this experiment for the equicorrelated and AR1 setting. We note that the FDR and power curves look very similar between the single and multiple knockoffs algorithm. If anything the results from the multiple knockoffs algorithm seem slightly favorable in the AR1 setting and with tighter distributions around the mean curves. However, we are cautious to draw such stronger conclusions since these results are purely empirical. It is reassuring to see that the multiple knockoffs approach does not empirically seem to perform any worse and controls the FDR in all simulation scenarios (also in the independent case, not shown).


\section{Clinical data application}
\label{sec:application}

Secukinumab (brand name Cosentyx), a human immunoglobulin G1-$\kappa$ monoclonal antibody that directly inhibits interleukin 17A, inhibits radiographic progression and has demonstrated long-term improvements in the signs and symptoms of patients with active Psoriatic Arthritis in several phase III trials \citep{MLpaper}. In this data application we will analyze a pool of four Phase III Cosentyx PsA trials, with the aim of identifying prognostic factors of a well established clinical outcome. Such factors are important as they may provide patients and their families with valuable information about the expected course of disease and might help guide in choosing an optimized dose. Further, they hold the promise of critical insights into the biology and natural history of the disease. Finally, prognostic factors are commonly used by drug developers to design, conduct, and analyze their clinical trials.

\subsection{Psoriatic Arthritis and ACR20}

Psoriatic arthritis (PsA) is a chronic inflammatory disease that affects peripheral and axial joints, entheses, and the skin, and is often
associated with impaired physical function and poor quality of life.
To measure improvement in patient's PsA conditions, the American College of Rheumatology \citep{ACR20} developed a criterion called $\mathrm{ACR20}$. This criterion is a composite binary response that assesses whether or not the patient experiences both improvement of 20\% in the number of tender and number of swollen joints, and a 20\% improvement in three of the following five criteria: patient global assessment, physician global assessment, functional ability measure [most often Health Assessment Questionnaire (HAQ)], visual analog pain scale, and high sensitivity C-reactive protein (hsCRP) or erythrocyte sedimentation rate. $\mathrm{ACR20}$ has become a common endpoint in PsA clinical trials and is often used as the primary endpoint to assess drug efficacy.

\subsection{Identifying prognostic factors of ACR20}

The underlying research problem involves identifying prognostic factors for the $\mathrm{ACR20}$ endpoint, measured at week $16$ after randomization, to different doses of Cosentyx (Placebo, 75 mg, 150 mg with or without loading and 300 mg). Patients with $\mathrm{ACR20}=1$ are defined as responders, while those with $\mathrm{ACR20}=0$ are defined as non-responders. The covariates included in the analysis are a mixture of patient demographics and disease characteristics as well as lab variables, measured at baseline (i.e. randomization). We seek to apply the knockoff methodology to discover which of these variables are prognostic of $\mathrm{ACR20}$ response.

\subsection{Data description and processing}

The data consists of four Phase III Cosentyx trials: FUTURE 2-5 \citep{future2, future3, future4, future5}, which total $2,148$ patients across five treatment arms (see \tabref{tab:trials}).

\begin{table}[ht]
	\centering
	\caption{Numbers of patients per trial (FUTURE2-5) and treatment arm. In parentheses we provide NCT numbers, unique identification codes given to each clinical study registered on ClinicalTrials.gov. NL stands for No Loading regimen.}
	\label{tab:trials}
	\begin{tabular}{@{}r|ccccc|l@{}}
        \toprule
		Trial \textbackslash Dose & Placebo & 75mg & 150mg NL & 150mg & 300mg & total \\
		\midrule
		FUTURE2 (NCT01752634) & 98 & 99 & 0 & 100 & 100 & 397 \\
		FUTURE3 (NCT01989468) & 137 & 0 & 0 & 138 & 139 & 414 \\
		FUTURE4 (NCT02294227) & 114 & 0 & 113 & 114 & 0 & 341 \\
		FUTURE5 (NCT02404350) & 332 & 0 & 222 & 220 & 222 & 996 \\
		\midrule
		total & 681 & 99 & 335 & 572 & 461 & 2,148 \\
        \bottomrule
	\end{tabular}
\end{table}

Before running the analysis we first apply standard data cleaning and processing procedures. In particular, we remove patients with missing data entries and identify pairs of highly correlated and clinically redundant variables (e.g. Body Mass Index (BMI) and Weight). For such redundant pairs we only pick one of the variables to include in the model. This ultimately results in a design matrix with no extreme collinearities and is in line with the work of \cite{Buhlmann_2013} that proposed to apply variable clustering and pick so-called cluster representatives prior to variable selection.

Finally, we recall that the sequential knockoff algorithm simulates the continuous columns of the knockoff matrix using a Gaussian distribution. Hence, to ensure that the Gaussian distributional assumption is reasonable we apply the normal score transformation on all continuous variables prior to running the knockoff analysis.

\subsection{Knockoff analysis} \label{subsec:knockoff_analysis}

Our data after pre-processing consists of complete observations for $n = 1,679$ patients and $p=58$ variables, of which $40$  are continuous and $18$ are factors (majority of them binary). In order to identify the prognostic factors of $\mathrm{ACR20}$ at week $16$, we repeatedly apply the sequential knockoff algorithm $1,000$ times. Each time we apply the sequential knockoff filter with a target FDR threshold of $q=0.20$.

\begin{figure}[tb]
	\centering
	\includegraphics[width=\textwidth]{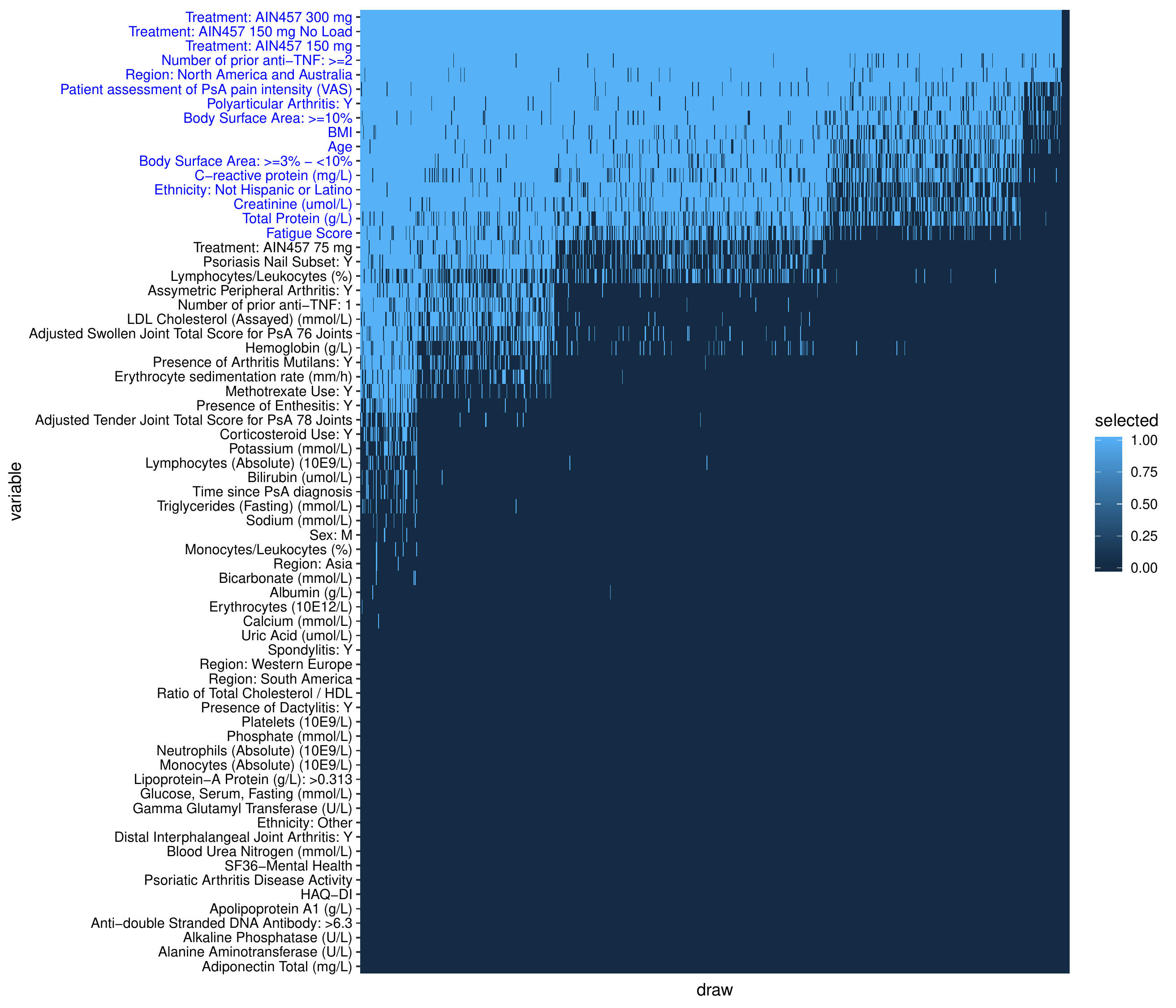}
	\caption{A heatmap demonstrating which variables were selected as prognostic of $\mathrm{ACR20}$ at week $16$ (light blue) across the $1,000$ runs of the sequential knockoff algorithm. A dark blue color indicates that a variable was not selected.}
	\label{fig:heatmap_acr20}
\end{figure}

In \figref{fig:heatmap_acr20} we see a heatmap of the most frequently selected variables across the $1,000$ knockoff draws. The variables highlighted in blue at the top of the list are deemed significant prognostic factors of $\mathrm{ACR20}$ at week $16$ according to the multiple knockoff criterion in \secref{subsec:multiknockoff}.

\begin{figure}[tb]
	\centering
	\includegraphics[width=\textwidth]{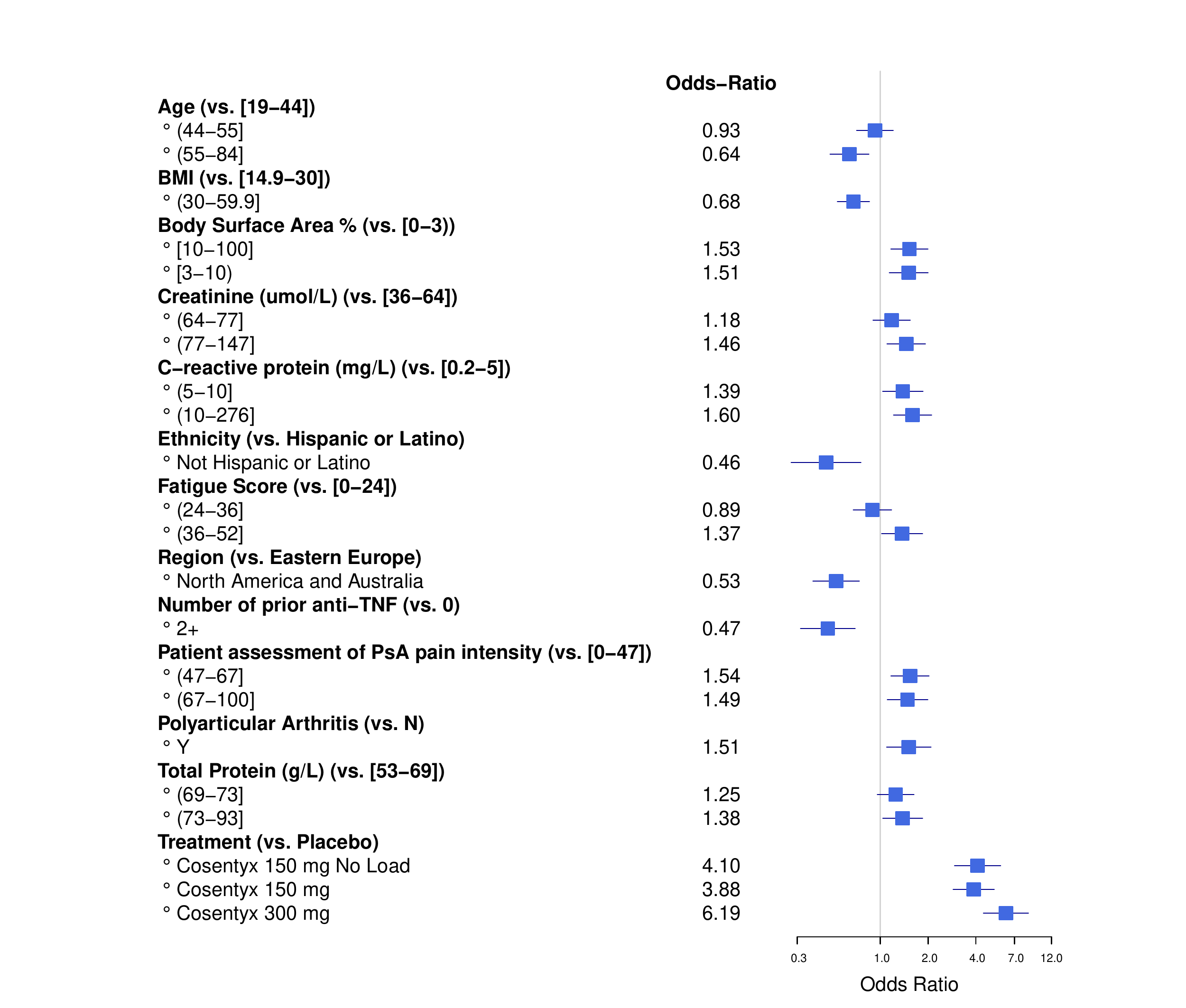}
	\caption{Forest plot of the odds ratios and corresponding 95\% confidence intervals of $\mathrm{ACR20}$ at week $16$ as reported from the logistic regression model involving the chosen variables from the multiple knockoff selection. For each factor variable (bold-faced) the reference level is displayed in parentheses.}
	\label{fig:forestplot_acr20}
\end{figure}

To validate clinical relevance, the selected set of covariates is fed into a standard logistic regression model with the binary $\mathrm{ACR20}$ endpoint as response. To ease clinical interpretation, we categorize continuous variables either based on terciles or well established clinical thresholds (if available). This allows us to report odds ratios of each factor level against a reference level; see Figure \ref{fig:forestplot_acr20}. It is somewhat reassuring that all the selected variables show prognostic effect, although ideally the model would need to be validated on an independent data set.

When interpreting the odds ratios, we note that the reported treatment arms greatly outperform the placebo arm ($\text{odds ratio}\approx4$ for the two $150$ mg doses and $\approx 6$ for the $300$ mg dose). This confirms the efficacy of the two regulatory approved doses of Cosentyx \citep{emalabel, fdalabel}.

Among the negative associations, we note that patients older than $55$ have lower odds of response than patients younger than $44$ and obese patients $(\mathrm{BMI}>30)$ have lower probability of response than non-obese $(\mathrm{BMI}\leq30)$. Patients who have had two or more prior anti-TNF treatments (prior to starting the trial) have lower odds of response than patients who have had no such prior treatment (this was previously described in studies and registries for patients switching treatment \citep{kavanaugh,glintborg}). In terms of demographics, patients in North America and Australia have lower odds of response than those in Eastern Europe and non-"hispanic or latino" patients have lower odds of response as compared to hispanic or latino. However, the prognostic effect of demographics is difficult to interpret directly, due to unmeasured socio-demographic differences and weak confounding with other prognostic factors such as obesity and varying access to treatment options for PsA prior to trial initiation.

Having any of the following symptoms at baseline: polyarticular arthritis, elevated pain intensity, or psoriatic skin condition (reported as part of the variable ``Body Surface Area'') increases patient's odds of response. Similarly, patients with hsC-reactive protein levels exceeding $5$mg/L (considered moderately high) or $10$mg/L (considered very high) have higher odds of response than those with normal levels (i.e. below $5$mg/L). hsC-reactive protein is a well known biomarker for systemic inflammation, seen in patients with PsA, and hence these results collectively indicate that the more active the disease is at baseline, the higher the probability of $\mathrm{ACR20}$ response over time. Conversely, having a high fatigue score ($\mathrm{FACIT}>36$), indicating less fatigue, has slightly higher odds of response as compared to patients with severe fatigue ($\mathrm{FACIT}<24$).

Two prognostic lab variables appeared on our list of discoveries, ``Total Protein'' and ``Creatinine''. The latter is a non-protein nitrogenous compound that is produced by the breakdown of creatine in muscle and is known to be associated with impaired renal function, which is a co-morbidity of several rheumatic diseases \citep{anders}. Both results indicate that elevated protein levels or their break-down products at baseline associate with higher odds of response. The significance of this finding is not immediately explicit, although a search in the literature has found significant associations between high expression of certain proteins in the serum and the rate of response in patients with psoriasis and arthritis \citep{heiskell,schett}.


\section{Conclusion}
\label{sec:conclusion}

In this paper we presented a practical sequential knockoff algorithm that may be applied to commonly encountered data sets involving both continuous and categorical (factor) covariates. We showed that this algorithm gives comparable performance to the MX knockoff algorithm \citep{knockoffpackage} when all covariates are Gaussian. Further, we demonstrated favorable performance as compared to competing methods in more general settings. In particular, the sequential knockoff algorithm consistently controlled FDR and had competitive power across a wide range of simulation scenarios. We also presented a heuristic multiple knockoffs algorithm, which empirically controlled FDR and had comparable power to the single knockoffs approach in our simulations. This method offers a practical procedure to assess robustness and stability of the (single) knockoff algorithm.
Finally, we applied the proposed methodology on a large clinical data pool consisting of $4$ studies of more than $2,000$  patients with Psoriatic Arthritis. The analysis revealed both known and novel prognostic factors of a well established clinical outcome.

One drawback of the sequential knockoff algorithm is that it is computationally expensive to run. On our computing system the sequential knockoff algorithm ran in a matter of minutes on a single simulated data set, while the conventional MX knockoff algorithm \citep{knockoffpackage} ran in a matter of seconds. This can certainly become an issue for larger data sets, but we believe that with a good High Performance Computing system the approach is practical for a wide range of applications. In this context, it is worth noting that the basic building block of the sequential knockoff algorithm, namely the elastic net routine with cross-validation, was not optimized in any way (e.g. by sharing a common penalty across steps of the algorithm). Hence there could still be room for some potential improvements and speed-ups.

In our simulations we noted that the power of the sequential knockoff method deteriorates as the inter-variable correlations become more extreme. This becomes intuitively clear when one considers the consequences of putting two perfectly correlated variables into the sequential knockoff algorithm. In this case the regression model in \algoref{alg:scip-approx} would lead to a perfect fit (for the collinear variables) and the resulting sequential knockoffs would perfectly match the original variables. To avoid extreme collinearity issues in our data analysis we searched for clinically redundant variables that were highly correlated (e.g. Body Mass Index and Weight) and picked a unique representative for such pairs. This is in line with the work of \cite{Buhlmann_2013} that proposed to apply variable clustering and pick so-called cluster representatives prior to variable selection. We believe that before analyzing any data set it's important to do some basic screening of variables to avoid feeding redundant information in the model. This was facilitated in our data application with domain knowledge of our clinical experts. However, in situations where the quantitative scientist has to handle hundreds or thousands of variables that are poorly understood (e.g. in genomic applications) automated screening methods may be required. \cite{Buhlmann_2013} provide interesting ideas in that direction and alternatively \cite{barber19} have proposed a lasso pre-screening step prior to running the knockoff algorithm. 

Finally, it's worth mentioning that in this paper we only discussed the sequential knockoff algorithm in the context of prognostic models. However, an obvious extension is to consider so-called predictive models that include an overall treatment effect $(T)$, additional main effects $(X)$, and finally treatment by main effect interactions $(X:T)$. For this scenario, one way forward could be to treat the treatment effect as fixed (i.e. always include it in the model) and simulate knockoffs of the main effects only $(\tilde{X})$. Since $\tilde{X}$ is a knockoff copy of $X$ it follows that $\tilde{X}:T$ is a knockoff copy of $X:T$. How one would then proceed to optimally determine both prognostic (significant main effects) and predictive (significant interactions) factors would require further investigation.

\subsection*{Acknowledgments}
We would like to thank Gregory Ligozio, Aimee Readie, Chris Holmes, and Thomas Nichols for helpful discussions during the course of this work.
\bibliographystyle{plainnat}%
\bibliography{knockoffs-paper}%

\appendix

\section{Algorithm description}
\label{sec:alg-desc}

\subsection{Estimating knockoff distribution - Sequential Elastic Net knockoff}
\label{app:senk}

In \secref{subsec:seqknockoff} we proposed a practical sequential knockoff algorithm for mixed-type data (\algoref{alg:scip-approx}). The algorithm involves estimating sequentially the distributions $ \cL(X_j \given X_{-j}, \tilde{X}_{1:j-1}) $, for $j=1,\dots,p$ and then simulating the corresponding knockoffs based on the estimated distributions. In particular, we model the above distributions with penalized linear (Gaussian) regression when $X_j$ is continuous and with penalized multinomial logistic regression when $X_j$ is categorical. We propose to use the elastic net \citep{zou05}, which has competitive prediction accuracy and can handle large number of covariates. The latter point is particularly important when the total number of predictors $p$ is large, but is convenient in any case since the number of covariates ($X_{-j}$ and $\tilde{X}_{1:j-1}$) grows as we loop through $j=1,\dots,p$. Most importantly, the elastic net has a fast and stable implementation for exponential dispersion models (in particular, linear and multinomial logistic regression) in the R-package \emph{glmnet} \citep{glmnet}.

Let $\eta_j = \eta(X_{-j}, \tilde{X}_{1:j-1};\beta)$ denote the linear predictor of the underlying exponential dispersion model, where $\beta$ denotes the regression coefficients linking the response $X_j$ with the covariates $X_{-j}$ and $\tilde{X}_{1:j-1}$. The objective function of the elastic net has the form
\begin{equation*}
\frac{1}{n} \sum_{i=1}^n \ell(X^{(i)}_j,\eta_j^{(i)}) + \lambda \left[(1-\alpha)\|\beta\|_2^2/2 + \alpha\|\beta\|_1 \right],
\end{equation*}
where $\ell(X^{(i)}_j,\eta_j^{(i)})$ denotes the negative log-likelihood contribution of observation $i$ (e.g. $\frac{1}{2}(X^{(i)}_j-\eta_j^{(i)})^2$ for the Gaussian case). The elastic net penalty is controlled by $\alpha \in [0,1]$ and is usually chosen by user (e.g. $\alpha=0$ for ridge regression and $\alpha=1$ for lasso regression). The overall penalty term $\lambda$ is estimated via cross-validation. According to simulations and real data analysis conducted in \cite{zou05} the elastic net generally outperforms lasso in prediction accuracy. However, our experimental results were not particularly sensitive to the exact choice of $\alpha$ and as a default we set $\alpha=1/2$.

Once we have an estimate of the linear predictor $\hat{\eta}_j$ we sample the knockoff
\begin{equation*}
\tilde{X}_j \sim \cN\left(\hat{\mu}=\hat{\eta}_j, \hat{\sigma}^2\right),
\end{equation*}
if $X_j$ is continuous, and
\begin{equation*}
\tilde{X}_j \sim {\mathrm{Multinom}}(\hat{\pi} =g^{-1}(\hat{\eta}_j))),
\end{equation*}
if $X_j$ is categorical (where $g(\cdot)$ denotes the generalized logit function of the multinomial logistic regression model). As for the estimation of the error variance $\hat{\sigma}^2$ in the Gaussian case we use the adjusted variance estimator $ \hat{\sigma}^2 = RSS/(n-\hat{s}) $, where $RSS$ is the Residual Sum of Squares and $ \hat{s} $ is the number of nonzero features selected from elastic net. This estimator has been shown to reduce estimation bias in finite samples \citep{stephen2016}.

\end{document}